\documentclass[prb, preprint]{revtex4}

\usepackage{graphicx}
\usepackage{bm}
\usepackage{dcolumn}
\usepackage{pifont}

\begin{document}

\title{
Field-induced spin reorientation in an antiferromagnetic Dirac material EuMnBi$_2$ revealed by neutron and resonant x-ray diffraction
}

\author{H. Masuda$^{1,2}$}
\author{H. Sakai$^{3,4}$}
\author{H. Takahashi$^{1,5}$}
\author{Y. Yamasaki$^{4,6,7}$}
\author{A. Nakao$^8$}
\author{T. Moyoshi$^8$}
\author{H. Nakao$^9$}
\author{Y. Murakami$^9$}
\author{T. Arima$^{7,10}$}
\author{S. Ishiwata$^{1,4,5}$}

\affiliation{
$^1$Department of Applied Physics, University of Tokyo, Tokyo 113-8656, Japan.\\
$^2$Institute for Materials Research, Tohoku University, Sendai 980-8577, Japan.\\
$^3$Department of Physics, Osaka University, Toyonaka, Osaka 560-0043, Japan.\\
$^4$PRESTO, Japan Science and Technology Agency, Kawaguchi, Saitama 332-0012, Japan.\\
$^5$Division of Materials Physics, Graduate School of Engineering Science, Osaka University, Toyonaka, Osaka 560-8531, Japan\\
$^6$Research and Services Division of Materials Data and Integrated System (MaDIS), National Institute for Materials Science (NIMS), Tsukuba 305-0047, Japan.\\
$^7$RIKEN Center for Emergent Matter Science (CEMS), Wako 351-0198, japan.\\
$^8$Comprehensive Research Organization for Science and Society, Tokai, 319-1106, Japan.\\
$^9$Condensed Matter Research Center and Photon Factory, Institute of Materials Structure Science, KEK, Tsukuba 305-0801, Japan.\\
$^{10}$Department of Advanced Materials Science, University of Tokyo, Kashiwa 277-8561, Japan.}

\date{}

\begin{abstract}
Field-dependent magnetic structure of a layered Dirac material EuMnBi$_2$ was investigated in detail by the single crystal neutron diffraction and the resonant x-ray magnetic diffraction techniques.
On the basis of the reflection conditions in the antiferromagnetic phase at zero field, the Eu moments were found to be ordered ferromagnetically within the $ab$ plane and stacked antiferromagnetically along the $c$ axis in the sequence of up-up-down-down.
Upon the spin-flop transition under the magnetic field parallel to the $c$ axis, the Eu moments are reoriented from the $c$ to the $a$ or $b$ directions forming two kinds of spin-flop domains, whereas the antiferromagnetic structure of the Mn sublattice remains intact as revealed by the quantitative analysis of the change in the neutron diffraction intensities.
The present study provides a concrete basis to discuss the dominant role of the Eu sublattice on the enhanced two-dimensionality of the Dirac fermion transport in EuMnBi$_2$.
\end{abstract}
\pacs{ }
\maketitle

\section{Introduction}
Dirac fermions in solids have attracted extensive attentions for their unusual transport properties.
The coupling between the Dirac fermion transport and the magnetism is of particular interest because of novel magnetotransport phenomena, as typified by the quantized anomalous Hall effect in magnetic topological insulator thin films\cite{QAHE_theory_Science2010, QAHE_exp_Science2013}.
Recently, a variety of magnetic Dirac or Weyl materials in bulk form have been reported, as exemplified by pyrochlore iridates\cite{iridate_PRB2011, iridate_AnnRev2014}, Mn$_3$Sn\cite{Mn3Sn_weyl_NatMat2017, Mn3Sn_weyl_NJP2017}, GdPtBi\cite{GdPtBi_Ong,GdPtBi_Suzuki}, EuTiO$_3$\cite{EuTiO3_weyl_SciAdv2018},  Co$_2$MnGa\cite{Co2MnGa_ARPES_Science2019}, and Co$_3$Sn$_2$S$_2$\cite{Co3Sn2S2_ARPES_Science2019, Co3Sn2S2_STM_Science2019}.
In these systems the magnetic order induces Weyl semimetal states, leading to the peculiar magnetotransport phenomena such as large anomalous Hall effects\cite{Mn3Sn_Nature2015, GdPtBi_Suzuki, EuTiO3_weyl_SciAdv2018, Sakai_Co2MnGa_Nernst_NatPhys2018, Co3Sn2S2_AHE_NatPhys2018, Co3Sn2S2_AHE_NatCommun2018} and negative magnetoresistances induced by the chiral anomalies\cite{Mn3Sn_weyl_NatMat2017, GdPtBi_Ong}.
In order to discuss the origin of these magnetotransport phenomena, it is indispensable to clarify the magnetic structure under external magnetic fields\cite{Eu2Ir2O7_resXray, R2Ir2O7_muon, Sm2Ir2O7_resXray, Nd2Ir2O7_neutron, Mn3Sn_neutron_JPSJ1982, Mn3Sn_neutron_JPCM1990, GdPtBi_AFM, GdPtBi_Suzuki, EuTiO3_neutron}.

\begin{figure}[tbp]
\begin{center}
\includegraphics[width=.6\linewidth]{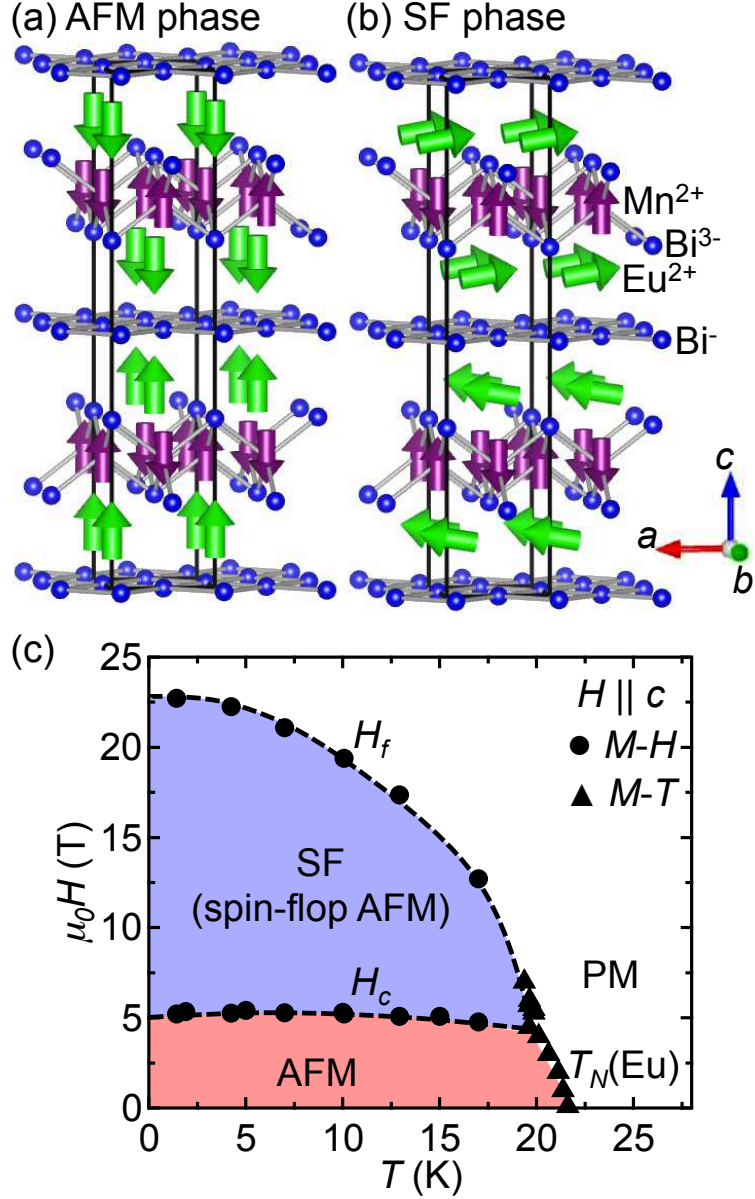}
\caption{
(a, b) Magnetic structures of EuMnBi$_2$ for the AFM and SF ($a-$domain) phases, respectively.
The magnetic structures were determined by the present work, while the atomic positions were reproduced from Ref. \onlinecite{May_EMB}.
The crystallographic unit cell is shown by the solid lines.
(c) Magnetic phase diagram of the Eu sublattice as functions of magnetic field ($H || c$) and temperature (reproduced from Ref. \onlinecite{Masuda_EMB}).
AFM, SF and PM denote Eu antiferromagnetic, spin-flop AFM and paramagnetic (Mn antiferromagnetic) phases, respectively.
Note that the Mn moments show antiferromagnetic order below $T_N({\rm Mn})\sim 315$ K. 
}
\label{fig:structure}
\end{center}
\end{figure}

Among the magnetic Dirac materials, EuMnBi$_2$ is a rare compound that exhibits distinct quantum transport of Dirac fermions coupled with the field-tunable magnetic order\cite{May_EMB, Masuda_EMB}.
EuMnBi$_2$ has a layered structure that consists of Bi square nets hosting two-dimensional Dirac fermions and the insulating layers hosting magnetic Eu$^{2+}$ and Mn$^{2+}$ ions as shown in Fig. \ref{fig:structure}(a, b)\cite{Park2011PRL_SMB, Lee2013PRB_SMB, Borisenko2015arXiv, May_EMB, Masuda_EMB}.
The magnetic phase diagram of the Eu sublattice under the external magnetic field $H$ parallel to the $c$ axis is shown in Fig. \ref{fig:structure}(c).
Upon the transition from the antiferromagnetic (AFM) to the spin-flop (SF) AFM phase phase, 
the interlayer resistivity $\rho_{zz}$ exhibits a sharp jump while the in-plane resistivity $\rho_{xx}$ remains almost the same, indicating the enhanced two-dimensionality in the SF phase\cite{Masuda_EMB}.
However, the mechanism of the coupling between the quantum transport of the Dirac fermion and the magnetic order in EuMnBi$_2$ remains unclear due to the lack of the detailed information on the magnetic structure.

Some of the authors previously reported a probable antiferromagnetic structure of the Eu sublattice for the AFM phase by the resonant x-ray magnetic diffraction techniques as shown in Fig. \ref{fig:structure}(a)\cite{Masuda_EMB}.
On the other hand, the magnetic arrangement of the Eu sublattice for the SF phase was not experimentally studied in detail.
Furthermore, the magnetic arrangement of the Mn sublattice below $T_N({\rm Mn})\sim315$ K was not studied\cite{May_EMB, Masuda_EMB}.
In this work, we have established the magnetic structures of both the Eu and Mn sublattices with particular focus on the SF phase, 
on the basis of the quantitative analysis of the single crystal neutron and resonant x-ray magnetic diffraction data under magnetic fields.

\section{Materials and Methods}
%Methods
Single crystals of EuMnBi$_2$ were grown by the Bi self-flux method\cite{Masuda_EMB, Canfield_flux}. 
EuMnBi$_2$ crystallizes in a tetragonal crystal structure with the space group of $I4/mmm$, $a=4.5416(4)$ \AA\  and $c=22.526(2)$ \AA\ as determined from the powder x-ray diffraction profile at room temperature \cite{May_EMB, Masuda_EMB}.
The crystal orientation was determined by x-ray Laue patterns.

Single crystal neutron diffraction experiments were carried out using the time-of-flight single-crystal neutron diffractometer SENJU at the Materials and Life Science Experimental Facility (MLF) of the Japan Proton Accelerator Research Complex (J-PARC).
The wavelength range of incident neutrons was selected to be $0.4 - 4.4$ \AA.
A plate-like single crystal sample with a dimension of $\sim3\times3.5\times1$ mm$^3$ was chosen for the experiments.
The neutron diffraction patterns in the magnetic field along the $c$ axis were measured using a vertical-field superconducting magnet
 for the AFM (2 K, 0 T), SF (2 K, 6 T) and PM (Eu paramagnetic and Mn antiferromagnetic, 25 K, 6 T) phases.

Resonant x-ray magnetic diffraction measurements near the Eu $L_3$ absorption edge ($E=$ 6.975 keV) were performed at BL-3A, Photon Factory, KEK, Japan\cite{Masuda_EMB}.
A single crystalline sample used for the measurements has a dimension of $\sim3\times2\times1.5$ mm$^3$ with the (1 0 $L$) ($L\sim1-2$) natural crystal facet.
The (4 0 1) reflection was measured at 5 K in the magnetic field along the $c$ axis using a vertical-field superconducting magnet equipped on a two-cycle diffractometer.
Polarization rotation analyses on the (4 0 1) reflection were performed using a Cu(110) single crystal.
Resonant x-ray magnetic diffraction measurements near the Eu $M_{4, 5}$ absorption edges ($E=$ 1.153, 1.125 keV) were performed at BL-19B, Photon Factory, KEK, Japan\cite{Nakao_BL19}.
A single crystalline sample used for the measurements has a dimension of $\sim2\times2\times1$ mm$^3$ with the (0 0 1) natural crystal facet.

\section{Result and Discussion}\label{sec:results}
\subsection{Neutron diffraction profiles}\label{sec:neutron}

\begin{figure*}[t]
\begin{center}
\includegraphics[width=.9\linewidth]{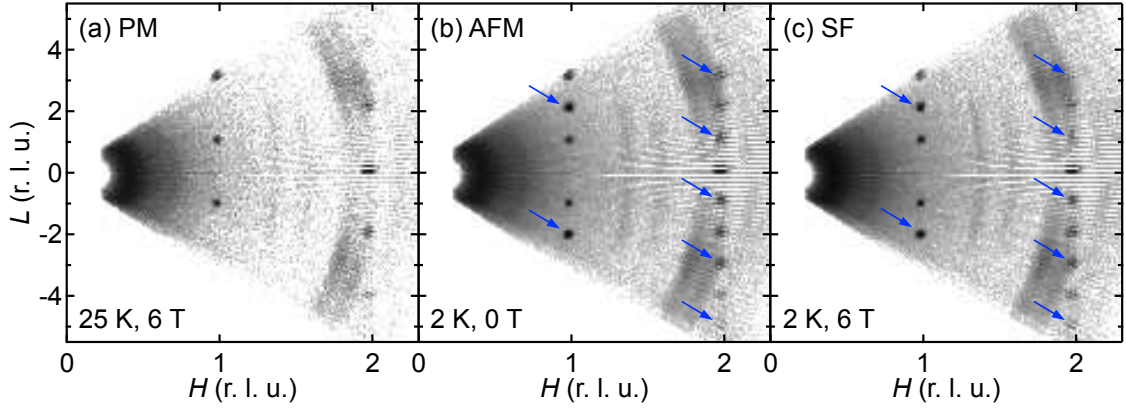}
\caption{
Neutron diffraction intensity distributions of EuMnBi$_2$ on the ($H$ 0 $L$) reciprocal lattice plane for the (a) PM (25 K, 6 T), (b) AFM (2 K, 0 T) and (c) SF (2 K, 6 T) phases, respectively.
The blue arrows in (b, c) indicate the magnetic reflections from the Eu sublattice that satisfy the conditions of $H+K+L=$ odd, $L\not=0$.
The ring-like intensities correspond to the powder lines which may arise from aluminum in the sample holder or bismuth flux stuck to the crystal surfaces.
}
\label{fig:ReciproMap}
\end{center}
\end{figure*}

Figures \ref{fig:ReciproMap}(a), (b) and (c) show the neutron diffraction intensity distributions on the ($H$ 0 $L$) reciprocal lattice plane in the PM (25 K, 6 T), AFM (2 K, 0 T) and SF (2 K, 6 T) phases, respectively.
The obtained lattice parameters at 2 K were $a=4.4988(2)$ \AA\ and $c=22.799(10)$ \AA.
%$a=4.49884\pm 0.00012/0.00021$ \AA and $c=22.7988 \pm 0.01011$ \AA %7941 AFM 15 LSUBmat
% $a=4.542$ \AA     and $c=22.52$ \AA     %7942 AFM 60 FindCell
In the PM phase, the observed reflections satisfy the condition of $H+K+L=$ even, which follows the extinction rule of the $I4/mmm$ symmetry of the crystal.
The antiferromagnetic arrangement of the Mn sublattice for the PM phase at 25 K far below $T_N$(Mn) $\simeq315$ K is derived as follows.
The absence of the magnetic reflections other than the ones superposed on the nuclear Bragg reflections indicates 
that the magnetic arrangement of the Mn sublattice is described by the propagation vector of ${\bm q}=(0,0,0)$.
Considering the reflection condition of $H+K+L=$ even in Fig. \ref{fig:ReciproMap}(a), the body-centered translational symmetry of the crystal with $I4/mmm$ is retained in the PM phase with the magnetic order of the Mn sublattice.
It follows that the Mn moments on two crystallographically equivalent sites related by the body-centered translation 
[{\it e.g.} (1/2, 0, 1/4) and (0, 1/2, 3/4); see Fig. \ref{fig:structure}(a)] are parallel to each other.
Furthermore, magnetization measurements imply that the Mn moments are aligned parallel to the $c$ axis below $T_N$(Mn)\cite{May_EMB}.
From these experimental facts, 
the magnetic structure of the Mn sublattice can be presumed to be 
antiferromagnetic for both in-plane and out-of-plane nearest neighbors with the spin direction along the $c$ axis [Fig. \ref{fig:structure}(a)], similar to isostructural SrMnBi$_2$\cite{Guo_SMB_neutron}.

In the AFM phase, magnetic superlattice reflections from the ordering of the Eu magnetic moments were observed at the positions of $H+K+L=$ odd, $L\not= 0$ as shown in Fig. \ref{fig:ReciproMap}(b).
This result is consistent with the previous results of the x-ray resonant magnetic diffraction measurements\cite{Masuda_EMB}.
The integer diffraction indices $HKL$ of the Eu magnetic reflections and the violation of the extinction rule of $H+K+L$ = even for the body-centered translation of the crystal with $I4/mmm$ reveal that the magnetic arrangement of the Eu sublattice is described by the propagation vector of ${\bm q}=(0,0,1)$ and the magnetic unit cell is the primitive tetragonal one.
Hence the magnetic moments on two crystallographically equivalent Eu sites related by the body-centered translation, {\it e.g.} (0, 0, $+z_0$) and (1/2, 1/2, $1/2+z_0$), $z_0\sim0.11$\cite{May_EMB}, are antiparallel to each other.
Furthermore, the absence of $L = 0$ Eu magnetic reflections indicates that two Eu sites facing across the Bi square net layer, {\it e.g.} $(0, 0, +z_0)$ and $(0, 0, -z_0)$, host Eu moments antiparallel to each other.
On the basis of these results, the magnetic structure of the Eu sublattice in the AFM phase can be determined as shown in Fig. \ref{fig:structure}(a), where the Eu moments order ferromagnetically within the $ab$ plane and align antiferromagnetically along the $c$ axis in the sequence of up-up-down-down\cite{Masuda_EMB}.

\begin{figure*}[tbp]
\begin{center}
\includegraphics[width=.8\linewidth]{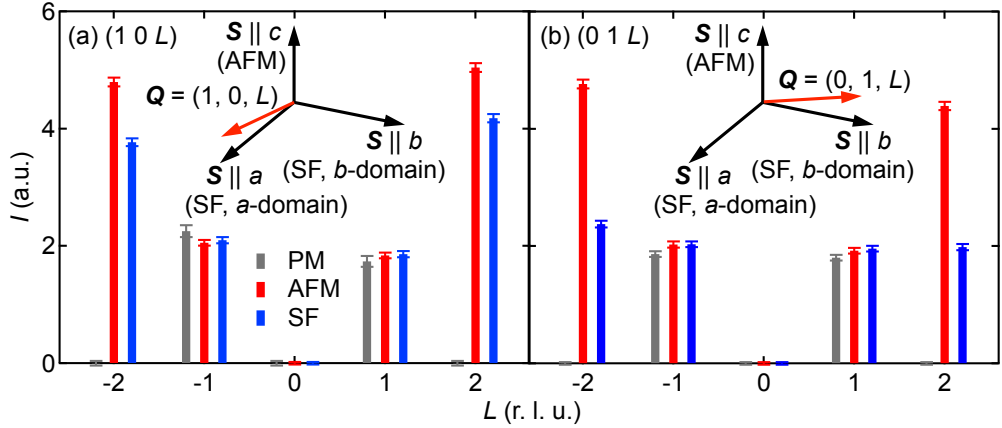}
\caption{
Integrated intensities of the (a) (1 0 $L$) and (b) (0 1 $L$) reflections ($-2\leq L\leq2$) for the AFM (2 K, 0 T), SF (2 K, 6 T) and PM (25 K, 6 T) phases.
Each inset show the schematic descriptions of the directions of the scattering vectors ${\bm Q}=(1, 0, L)$ and $(0, 1, L)$ ($L\simeq2$), along with the directions of the Eu moments ${\hat {\bm S}}$ in the AFM and SF phases.
}
\label{fig:IvsL}
\end{center}
\end{figure*}

The diffraction intensity distribution for the SF phase is qualitatively similar to that for the AFM phase as shown in Fig. \ref{fig:ReciproMap}(c), suggesting that the magnetic arrangement for the SF phase is similar to that for the AFM phase except for the orientations of the magnetic moments.
Figures \ref{fig:IvsL}(a) and (b) show the integrated intensities of the (1 0 $L$) and (0 1 $L$) reflections ($-2\leq L\leq 2$), respectively, in the PM, AFM and SF phases.
Reflections of $L=$ odd, {\it i.e.} $H+K+L=$ even, arising from the nuclear and Mn magnetic diffractions show comparable intensities in the PM, AFM and SF phases.
This result implies that the Mn moments show similar magnetic structures in three phases, which will be discussed more quantitatively in Sec. \ref{sec:quantitative}.
On the other hand, intensities of the (1 0 $\pm2$) and (0 1 $\pm 2$) reflections arising from the Eu magnetic order for the SF phase are significantly smaller than that for the AFM phase.
This result is interpreted in terms of the reorientation of the Eu moments from the $c$ direction to the $a$ or $b$ directions upon the transition from the AFM phase to the SF phase.
The neutron magnetic diffraction intensities arise from the component of the magnetic moments perpendicular to the scattering vector ${\bm Q}$.
Since ${\bm Q}=(1,0,\pm2)$ is nearly parallel to the $a$ axis as shown in the inset of Fig. \ref{fig:IvsL}(a), the (1 0 $\pm2$) magnetic reflections mostly arise from the $c$ and $b$ component of the Eu moments.
Likewise, the (0 1 $\pm2$) reflections arise from the $c$ and $a$ component of the Eu moments (inset to Fig. \ref{fig:IvsL}[b]).
Therefore the larger intensities of the (1 0 $\pm2$) and (0 1 $\pm2$) Eu magnetic reflections for the AFM phase than SF phase indicate that the Eu moments are aligned parallel to the $c$ axis in the AFM phase, while they are oriented parallel to the $ab$ plane.
Furthermore, in the SF phase, intensities of the (1 0 $\pm2$) magnetic reflections are larger than those of the (0 1 $\pm2$) reflections as shown in Figs. \ref{fig:IvsL}(a) and (b).
This result indicates that the major number of Eu moments are oriented along the $b$ direction and the others are oriented along the $a$ direction in the SF phase.
This implies that two types of domains exist in the SF phase where Eu moments are aligned parallel to the $a$ and $b$ axes (mentioned as $a-$ and $b-$domains in the following), and the $b-$domain is somewhat dominant.
The $b-$domain is favored presumably due to the small misalignment of the magnetic field away from the $c$ axis\cite{}.
Quantitative estimate of the domain volume fraction is given in Sec. \ref{sec:quantitative}.

\subsection{Resonant x-ray magnetic diffraction}\label{sec:XRMS}
\begin{figure}[tbp]
\begin{center}
\includegraphics[width=.7\linewidth]{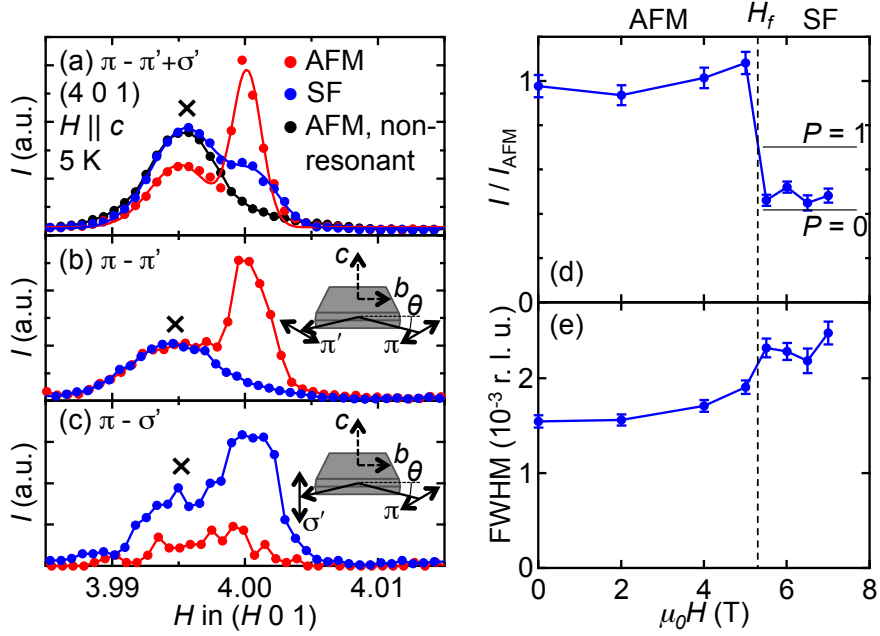}
\caption{
(a) Peak profiles of the (4 0 1) resonant x-ray magnetic reflection along [1 0 0] at Eu $L_3$ edge ($E=6.975$ keV) for the AFM (5 K, 0 T) and SF (5 K, 7 T) phases\cite{Masuda_EMB}. 
Peak profile at non-resonance ($E = 7$ keV) for the AFM phase is also shown.
The broad peak denoted by $\times$ arises from an unknown powder line.
(b, c) Peak profiles of the (4 0 1) resonant x-ray magnetic reflection in the (b) $\pi-\pi'$ and (c) $\pi-\sigma'$ channels, respectively, for the AFM and SF phases.
Schematic configurations for the measurements are shown as insets. 
$\theta\simeq52^\circ$ is the scattering angle. 
$\alpha\simeq3^\circ$, the angle between the $ab$ plane and the scattering plane, is not shown.
(d, e) Magnetic field dependence of the normalized intensity and the full width at half maximum (FWHM) along [1 0 0] of the (4 0 1) resonant x-ray magnetic reflection. 
$I_{\rm AFM}$ is the averaged intensity for the AFM phase at 5 K.
The vertical dashed line denotes $H_f\simeq5.3$ T, the transition field from the AFM phase to the SF phase\cite{Masuda_EMB}.
The horizontal lines in (d) indicate the ratios of the intensities for the SF and AFM phases $I_{\rm SF}/I_{\rm AFM}$ calculated using Eq. \ref{eq:XRMSint} for fully polarized spin-flop domains ($P=0$ or $P=1$).
}
\label{fig:XRMS}
\end{center}
\end{figure}

The difference in orientations of the Eu moments between the AFM and SF phases has also been signified by the resonant x-ray magnetic diffraction measurements.
We observed the resonant x-ray magnetic diffraction from the Eu sublattice near the Eu $M_{4, 5}$ ($E=$1.153, 1.125 keV) and $L_3$ (6.975 keV) edges in the AFM phase (see Fig. S1 in the supplementary material).
Here we focus on the (4 0 1) magnetic reflection at the Eu $L_3$ edge ($E= 6.975$ keV) as shown in Fig. \ref{fig:XRMS}(a)\cite{Masuda_EMB}.
Although the (4 0 1) magnetic reflection was observed both in the AFM and SF phases, the intensity in the latter phase is much smaller than that in the former, indicating the reorientation of the Eu moments.

In order to determine the orientation of the Eu moments, we performed the polarization analysis for the magnetic reflection.
The peak profile of the (4 0 1) magnetic reflection in the $\pi-\pi'$ channel is shown in Fig. \ref{fig:XRMS}(b), 
where the reflection was observed only for the AFM phase.
On the other hand, the magnetic reflection in the $\pi-\sigma'$ channel was observed only for the SF phase as shown in Fig. \ref{fig:XRMS}(c).
Since the resonant x-ray magnetic reflection in the electric-dipole transition arises from the component of the magnetic moment perpendicular to both incident and scattered polarization vectors\cite{Hannon1988PRL_XRMS}, 
the magnetic reflection in the $\pi-\pi'$ channel arises from the component of the magnetic moment nearly parallel to the $c$ axis as seen from the inset to Fig. \ref{fig:XRMS}(b).
Therefore the observation of the magnetic reflection in the $\pi-\pi'$ channel for the AFM phase indicates that the Eu moments are aligned parallel to the $c$ axis.
Likewise, the magnetic reflection in the $\pi-\sigma'$ channel arises from the component of the magnetic moment perpendicular to the $c$ axis (inset to Fig. \ref{fig:XRMS}[c]), 
hence the observation of the magnetic reflection in the $\pi-\sigma'$ channel for the SF phase indicates that the Eu moments are aligned parallel to the $ab$ plane.

Figures \ref{fig:XRMS}(d) and (e) show the magnetic field dependence of the intensity and the FWHM along [1 0 0] of the (4 0 1) resonant x-ray magnetic reflection, respectively.
The reflection intensity shows a sharp drop at $H_f\sim5.3$ T, reflecting the spin-flop transition.
The ratio of the averaged intensities for the AFM phase ($H=$ 0, 2, 4, 5 T) and that for the SF phase ($H=$ 5.5, 6, 6.5, 7 T) was $I_{\rm SF}/I_{\rm AFM}=0.482(19)$.
This ratio is calculated as 
\begin{eqnarray}
\frac{I_{\rm SF}}{I_{\rm AFM}} 
&=&P\frac{4\sin^2\theta\cos^2\theta\sin^2\alpha+\sin^2\theta\cos^2\alpha}{4\sin^2\theta\cos^2\theta\cos^2\alpha+\sin^2\theta\sin^2\alpha}\nonumber\\
&+&\left(1-P\right)\frac{\cos^2\theta}{4\sin^2\theta\cos^2\theta\cos^2\alpha+\sin^2\theta\sin^2\alpha}%\nonumber\\
\label{eq:XRMSint}
\end{eqnarray}
Here, $P$ is the volume fraction of the $a-$domain, $\theta\simeq52^\circ$ is the scattering angle and $\alpha\simeq3^\circ$ is the angle between the scattering plane and the $ab$ plane\cite{Hannon1988PRL_XRMS}.
Eu moments are assumed to be aligned parallel to the $c$ axis for the AFM phase and to the $a$ and $b$ axes for the SF phase in the $a-$ and $b-$domains, respectively.
Note here that the $\frac{I_{\rm SF}}{I_{\rm AFM}}$ values calculated for fully polarized spin-flop domains ($P=0$ and 1) are indicated by the horizontal lines in Fig. \ref{fig:XRMS}(d).
From the experimental  $\frac{I_{\rm SF}}{I_{\rm AFM}}$ value, the domain ratio is estimated to be $P=0.31(8)$.
The FWHM of the (4 0 1) magnetic reflection slightly increases above $H_f\sim5.3$ T possibly due to the formation of the spin-flop domains in the SF phase.

\subsection{Quantitative analysis on the neutron diffraction data}\label{sec:quantitative}
So far, we have qualitatively discussed the magnetic structures of the Mn sublattice for the PM, AFM and SF phases and that of the Eu sublattice for the AFM and SF phases.
Here, we present a quantitative analysis on the relative neutron diffraction intensities for the SF and AFM phases with a particular focus on the impact of the Eu spin-flop on the magnetic structure of the Mn sublattice.

\begin{figure}[tbp]
\begin{center}
\includegraphics[width=.7\linewidth]{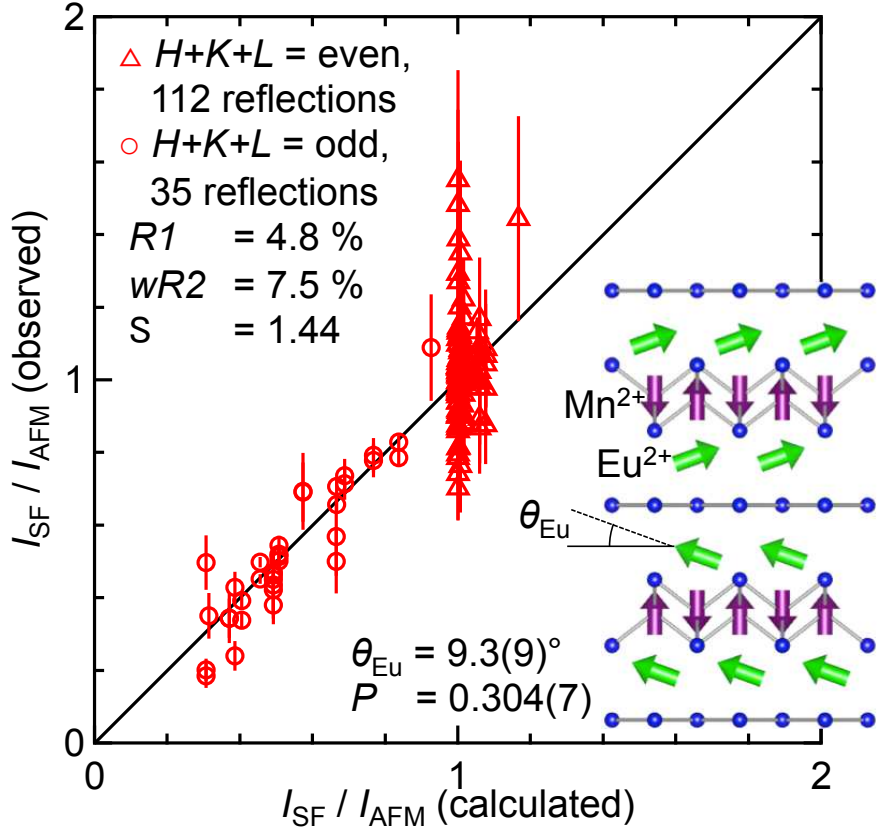}
\caption{
Comparison between the observed and calculated ratio of the intensities for the SF and AFM phases in EuMnBi$_2$.
Inset depicts the magnetic structure for the SF phase, along with the definition of $\theta_{\rm Eu}$.
}
\label{fig:fitting}
\end{center}
\end{figure}
The neutron diffraction intensities were collected for several crystal orientations in the PM, AFM and SF phases.
The observed ratios of the intensities in the SF and AFM phases $I_{\rm SF}/I_{\rm AFM}$ are plotted against the calculated ones in Fig. \ref{fig:fitting}.
For the calculation, the magnetic structure for the AFM phase was fixed to that shown in Fig. \ref{fig:structure}(a) based on the discussions in Sec. \ref{sec:neutron}.
For the SF phase, the magnetic structure shown in the inset of Fig. \ref{fig:fitting} was assumed, where the Eu moments are canted by $\theta_{\rm Eu}$ from the $ab$ plane to the direction of the magnetic field.
As we discussed in Sec. \ref{sec:neutron}, the magnetic structure of the Mn sublattice for the SF phase is not significantly different from those for the PM and AFM phases.
Therefore, the magnetic structure of the Mn sublattice for the SF phase was first fixed to be the same as that for the AFM phase.
The crystal structure parameters were also fixed to the reported values\cite{May_EMB}.
The amplitude of the Eu magnetic moment $M_{\rm Eu}$ was fixed to 6.4 $\mu_{\rm B}$, which is the saturation magnetization of EuMnBi$_2$ at 1.4 K\cite{Masuda_EMB}.
The amplitude of the Mn magnetic moment $M_{\rm Mn}$ was also fixed to 3.4 $\mu_{\rm B}$, 
which is taken from the experimental value for isostructural SrMnBi$_2$\cite{Guo_SMB_neutron}.
The magnetic form factors of Eu and Mn were taken from Ref. \onlinecite{MagFormFact}.
Two parameters, $P$ and $\theta_{\rm Eu}$, were refined using 
112 reflections under the condition of $H+K+L=$ even and 35 reflections under the condition of $H+K+L=$ odd under the conditions of $\sin\theta/\lambda<0.5$ \AA$^{-1}$ and $I>5\sigma$\cite{supplement2}.

As shown in Fig. \ref{fig:fitting}, the observed and calculated intensities $I_{\rm SF}/I_{\rm AFM}$ agree with each other with the reasonable reliable factors $R1=4.8$ \%, $wR2=7.5$ \% and the goodness of fit $S=1.44$.
The refined domain volume fraction $P=0.304(7)$ indicates sufficient dominance of the $b-$domain over the $a-$domain in the SF phase, 
which is likely due to the misalignment of the magnetic field away from the $c$ axis.
The refined canted angle of the Eu moments $\theta_{\rm Eu}=9.3(9)^\circ$ indicates the net magnetization of $M_{\rm Eu}\sin\theta_{\rm Eu}=1.05(10)\ \mu_{\rm B}$/Eu, which is comparable to the magnetization of 1.6 $\mu_{\rm B}$/Eu observed at 1.4 K, 6 T\cite{Masuda_EMB}.

It should be noted here that the intensity ratios for the SF and AFM phases $I_{\rm SF}/I_{\rm AFM}$, rather than the intensities themselves, were used for the quantitative analysis in order to avoid the effect of strong neutron absorption by Eu.
Since the neutron absorption cross section is independent of the external magnetic field and the magnetic structure,
$I_{\rm SF}/I_{\rm AFM}$ is in principle unaffected by the neutron absorption when $I_{\rm AFM}({\bm Q})$ and $I_{\rm SF}({\bm Q})$ were measured in the same crystal configuration.
Note that the large variation and error bar of the experimental $I_{\rm SF}/I_{\rm AFM}$ values around $I_{\rm SF}/I_{\rm AFM}=1$ in Fig. \ref{fig:fitting} stem from the relatively weak intensity (see Fig. S3 in the supplementary material).

We further proceeded our analysis by assuming the magnetic structure for the SF phase, where the Mn moments are canted to the in-plane direction due to the interaction between the Eu and Mn moments (see Fig. S4 in the supplementary material).
However, the agreement between the observed and calculated $I_{\rm SF}/I_{\rm AFM}$ was not improved.
This result shows that the magnetic structure of the Mn moments in the SF phase is the same as that in the AFM phase within the experimental accuracy. 

\subsection{Role of magnetism on the magnetoresistance effect}
Finally, we briefly discuss the origin of the enhanced two-dimensionality in the SF phase in EuMnBi$_2$. Upon the spin-flop transition, the Eu magnetic moments are reoriented from the $c$ direction to the $a$ or $b$ directions, while keeping the same antiferromagnetic arrangement.
Considering the experimental fact the magnetic structure of the Mn moments is virtually unchanged upon the spin-flop transition, 
it is reasonable to presume that the Mn moments play negligible role on the magnetoresistance effects of EuMnBi$_2$.
Therefore, we focus on the role of the reorientation of the Eu moments upon the spin-flop transition.

First we consider the possible effect of magnetic domain walls between the two spin-flop domains on the enhanced two-dimensionality in the SF phase.
However, this possibility can be ruled out by the magnetoresistance measurements under tilted magnetic fields.
In fact, the increase in $\rho_{zz}$ upon the spin-flop transition can be observed when the magnetic field is tilted away from the $c$ axis by 65$^\circ$\cite{Masuda_EMB_LLsplitting}, 
where the spin-flop domains are expected to be disappeared.

Next, we consider the possible effect of the reorientation of the Eu moments on the band dispersion along the $k_z$ ($c^\ast$) direction.
It should be noted here that the energy scale of the transfer between the Bi conduction layers via the local Eu moments is expected to be unchanged upon the spin-flop transition, since the orbital of Eu$^{2+}$ high-spin state ($S=7/2$, $L=0$) is essentially isotropic\cite{Masuda_EMB}.
Here we point out the experimental fact that the magnetic unit cell in the AFM and SF phases is the primitive tetragonal cell with two Bi square net layers (Fig. \ref{fig:structure}[a]), which would fold the Dirac band along the $k_z$ direction to form two quasi-2D Dirac bands\cite{Masuda_EMB_LLsplitting}.
The gap between the two Dirac bands at the zone boundary would suppress the $k_z$ dispersion, which is consistent with the increase in $\rho_{zz}$ upon the transition from the PM to the AFM phase\cite{Masuda_EMB}.
The reorientation of the Eu moment upon the spin-flop transition breaks the 4-fold rotational symmetry, which allows additional mixing between Bi $p_x$, $p_y$ orbitals through the spin-orbit coupling,
which may enhance the zone boundary gap and further suppress the $k_z$ dispersion of the two Dirac bands.
While this may account for the enhanced two-dimensionality in the SF phase, more experimental and theoretical studies would be necessary to support this possibility.

\section{Conclusions}
We have established the magnetic structure of EuMnBi$_2$ under magnetic field up to 6 T by the single crystal neutron diffraction and the resonant x-ray magnetic diffraction techniques.
In the AFM phase below $T_N$(Eu) $\simeq22$ K, the Eu moments are ordered ferromagnetically within the $ab$ plane with the moments aligned along the $c$ axis, 
which are stacked antiferromagnetically along the $c$ axis.
The Eu moments are reoriented to the $a$ or $b$ directions forming the domains upon the spin-flop transition to the SF phase under the magnetic fields along the $c$ axis.
The Mn sublattice with the checkerboard-type AFM order is apparently less affected by the reorientation of the Eu moments.
These results offer a concrete basis to discuss the role of the Eu magnetic order on the two-dimensionality of the Dirac fermions on the Bi layers in EuMnBi$_2$.

This study was supported in part by KAKENHI (Grant Nos. 16H06015, 17H01195, 19H01851), JST PRESTO Hyper-nano-space Design toward Innovative Functionality (Grant No. JPMJPR1412), JST PRESTO Scientific Innovation for Energy Harvesting Technology (No. JPMJPR16R2) and the Asahi Glass Foundation.
The neutron diffraction experiment on SENJU (BL-18) at MLF J-PARC was performed under the approval of the Proposal No. 2016B0151.
The resonant x-ray diffraction experiment on BL-3A and BL-19B at PF KEK was performed under the approval of the Proposal Nos. 2012S2-005, 2015S2-007 and 2016PF-BL-19B.


\begin{thebibliography}{100}
\bibitem{supplement2} Reflections in the range of $\sin\theta/\lambda>0.5$ \AA$^{-1}$ were not used for the analysis because they are less informative for the magnetic structure. See Fig. S2 for more detail.
\end{thebibliography}
\end{document}